\documentstyle[emulateapj,danonecolfloat]{article}
\def\NoApjSectionMarkInTitle#1{#1.\ }

\def\beq#1{\begin{equation}\label{#1}}
\def\eeq  {\end{equation}}
\def\beqa#1{\begin{eqnarray}\label{#1}}
\def\eeqa  {\end{eqnarray}}
\def\eq#1 {equation~(\ref{#1})}
\def\Eq#1 {Equation~(\ref{#1})}

\def\Cbf {{\bf C}}
\def\dT  {\delta T}
\def\ha  {$H_{\alpha}$~}
\def\haa {$H_{\alpha}$}
\def\l   {\ell}
\def\nbf {{\bf n}}

\def\SS  {{\bf\Sigma}}
\def\xbf {{\bf x}}
\def\Xbf {{\bf X}}

\def\Xbft{{\bf X}^T}
\def\ybf {{\bf y}}
\def\albf{{\bf a}}
\def\albfHat{\widehat{{\bf a}}} 
\def\da     {\delta\albfHat}

\def\etal   {{\it et al.~}}
\def\lletter{$Letter$}
\def\microk {$\mu$K~}
\def\mmicrok{$\mu$K}
\def\microm {$\mu$m~}
\def\mmicrom{$\mu$m}
\def\mj     {MJy/sr}
\def\syn    {synchrotron~}

\def\um     {\mu{\rm m}}

\def\ie{{\frenchspacing\it i.e.}}
\def\eg{{\frenchspacing\it e.g.}}

\def\ith{i^{th}}
\def\jth{j^{th}}
\def\expec#1{\langle#1\rangle}

\def\spose#1{\hbox to 0pt{#1\hss}}
\def\simlt{\mathrel{\spose{\lower 3pt\hbox{$\mathchar"218$}}
    \raise 2.0pt\hbox{$\mathchar"13C$}}}
\def\simgt{\mathrel{\spose{\lower 3pt\hbox{$\mathchar"218$}}
    \raise 2.0pt\hbox{$\mathchar"13E$}}}
\def\simpropto{\mathrel{\spose{\lower 3pt\hbox{$\mathchar"218$}}
    \raise 2.0pt\hbox{$\propto$}}}

\def\fig#1{Figure~\ref{#1}}
\def\Fig#1{Figure~\ref{#1}}

\def\rn{\noindent\parshape 2 0truecm 8.8truecm 0.3truecm 8.5truecm}
\def\nn#1 #2{#1, #2.}				
\def\nnn#1 #2 #3{#1, #2. #3.}			
\def\nnnn#1 #2 #3 #4{#1, #2. #3. #4.}		
\def\nnnnn#1 #2 #3 #4 #5{#1, #2. #3. #4. #5.}	
\def\rg#1;#2;#3;#4;#5;#6 {\par\rn#1 #2, {\it #3}, {\bf #4}, #5 (``#6'') \par}
\def\rf#1;#2;#3;#4;#5 {\par\rn#1 #2, {\it #3}, {\bf #4}, #5\par}
\def\rfbook#1;#2;#3;#4;#5 {{\frenchspacing\par\rn#1 #2, {\it #3} (#4: #5)\par}}
\def\rfproc#1;#2;#3;#4;#5;#6 {{\frenchspacing\par\rn#1 #2, in {\it #3}, ed. #4 (#5: #6)\par}}
\def\rfprep#1;#2;#3  {{\par\rn#1 #2, #3\par}}
\def\rfprepp#1;#2;#3 {{\par\rn#1 #2, #3\par}}

\journalid{337}{15 January 1989}
\articleid{11}{14}
\slugcomment{\today; ApJ, in press}

\begin{document}
\twocolumn[

\title{A New Spin on Galactic Dust}

\author{Ang\'elica de Oliveira-Costa\footnote{Department of Physics \& Astronomy, 
                                              University of Pennsylvania, 
					      Philadelphia, PA 19104, USA}, 
                                   Max Tegmark$^{a}$,
	       Douglas P. Finkbeiner\footnote{Department of Astronomy, 
	       				      Princeton University, 
					      Princeton, NJ 08544, USA}, \\			      
			 R.D. Davies\footnote{University of Manchester, 
	                		      Nuffield Radio Astronomy Laboratories, 
					      Jodrell Bank, 
					      Cheshire, SK11 9DL, UK},
		 Carlos M. Gutierrez\footnote{Instituto de Astrofisica de Canarias, 
	                        	      38200 La Laguna, 
					      Tenerife, Spain},
        		L.M. Haffner\footnote{Astronomy Department, 
	                		      University of Wisconsin, 
					      Madison, WI 53706, USA},
		       Aled W. Jones\footnote{Mullard Radio Astronomy Observatory, 
	                		      Cavendish Laboratory, 
					      Cambridge, CB3 0HE, UK},
				  A.N. Lasenby$^{f}$, \\ 
				     R. Rebolo$^{d}$,  
        		       Ron J. Reynolds$^{e}$,
        		  S.L. Tufte\footnote{Physics Department, 
	                		      Lewis and Clark College, 
					      Portland, OR 97219, USA} and 
	                           R.A. Watson$^{c,d}$} 



\begin{abstract}
We present a new puzzle involving Galactic microwave emission and 
attempt to resolve it. On one hand, a cross-correlation analysis of 
the WHAM H$\alpha$ map with the Tenerife 10 and 15 GHz maps shows that 
the well-known DIRBE correlated microwave emission cannot be dominated 
by free-free emission. On the other hand, recent high resolution observations 
in the 8--10 GHz range with the Green Bank 140 ft telescope by Finkbeiner \etal 
failed to find the corresponding $8\sigma$ signal that would be expected in 
the simplest spinning dust models. So what physical mechanism is causing this 
ubiquitous dust-correlated emission? We argue for a model predicting that 
spinning dust is the culprit after all, but that the corresponding small 
grains are well correlated with the larger grains seen at 100\microm only on 
large angular scales. 
In support of this grain segregation model, we find the best spinning dust 
template to involve higher frequency maps in the range 12-60\mmicrom, where emission 
from transiently heated small grains is important. Upcoming CMB experiments such as 
   ground--based interferometers, MAP and Planck LFI
with high resolution at low frequencies should allow a definitive test of this 
model.
\end{abstract}

]


\section{INTRODUCTION}

Understanding the physical origin of Galactic microwave emission 
is interesting for two reasons: to determine the fundamental 
properties of the Galactic components, and to refine the modeling 
of foreground emission in Cosmic Microwave Background (CMB) experiments. 
There are three Galactic foregrounds currently identified:
synchrotron radiation, free-free emission and thermal (vibrational) 
emission from dust grains. In the last few years, however, it has become 
clear that a fourth component exists. This component, which we nickname  
``Foreground X'', is spatially correlated with 100\microm dust emission 
but with a spectrum rising towards lower frequencies in a manner that is 
incompatible with thermal dust emission.  

This fourth component was first discovered in the COBE DMR data by 
Kogut \etal (1996a; 1996b), who tentatively identified it as free-free 
emission. It has since been detected in the data sets from 
	Saskatoon         (de Oliveira-Costa \etal 1997), 
	OVRO              (Leitch \etal 1997), 
	the 19 GHz survey (de Oliveira-Costa \etal 1998) and
	Tenerife          (de Oliveira-Costa \etal 1999, hereafter dOC99; 
			   Mukherjee \etal 2000). 
Draine and Lazarian (1998, hereafter DL98) argued against the free-free hypothesis
on energetic grounds, and suggested that Foreground X was caused by 
dust after all, but via rotational rather that vibrational emission.
The fact that the spectrum of Foreground X is observed to flatten towards 
lower frequencies and perhaps even turn over around 15 GHz
(dOC99) agrees well with the spinning dust predictions\footnote{
		Adding information to this puzzle, correlations 
		between dust and \ha maps are marginal (McCullough 1997; 
		Kogut 1997).
		}. 
Although a consensus has still not been reached on this point 
(Mukherjee \etal 2000), the case for spinning dust was beginning 
to look quite solid until recently.

New observations done by Finkbeiner \etal (2001, hereafter F01) have 
reopened the question about the existence and nature of Foreground X. 
Using the Green Bank 140 ft telescope, these authors observed 10 IRAS 
dust filaments at arcminute (FWHM$\sim$6') scales in the frequencies 
5, 8 and 10~GHz\footnote{  
	The selected regions were carefully chosen to be cold neutral 
	clouds or HII regions at low density because in both objects
	the free-free emission is expected to be subdominant.}. 
Although the Draine and Lazarian spinning dust model normalized to the 
Tenerife observations suggested that several filaments should be detected 
at the $8\sigma$ level, F01 observations shows that only two of these 10 
regions are correlated with the 100\microm dust map, and only one of these 
two detections is compatible with the Tenerife results. Moreover, their
frequency spectra is consistent with spining dust model of DL98 and 
inconsistent with free-free emission alone. 

\begin{figure*}[tb]
\vskip-3.0cm
\centerline{{\vbox{\epsfxsize=17.2cm\epsfbox{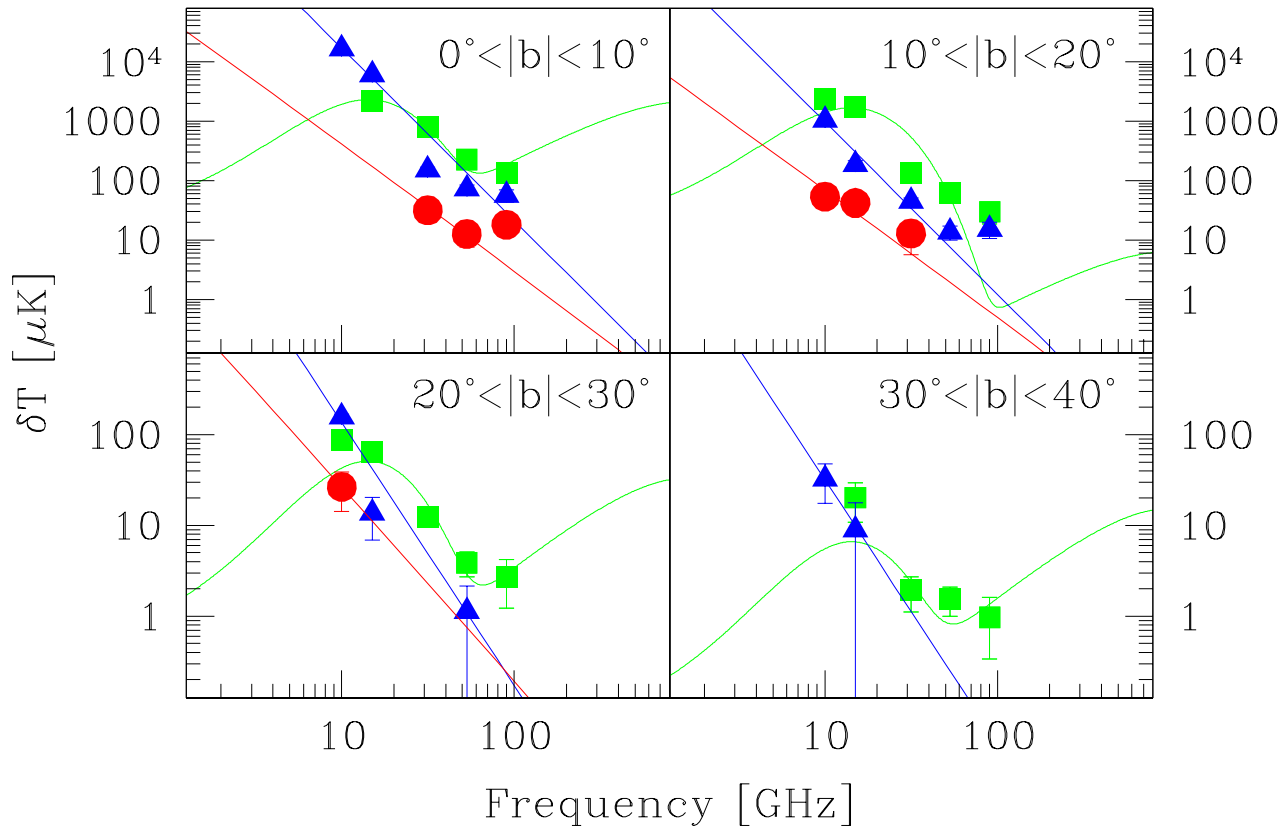}}}}
\vskip-2.6cm
\figcaption{\label{dTFig} 
    The frequency dependence of rms foreground fluctuations 
    is shown for four different Galactic latitude slices. 
    Squares show  DIRBE$-$correlated emission (Foreground X),
    triangles    Haslam$-$correlated emission (synchrotron), and
    circles        \haa$-$correlated emission (free-free) 
    for four Galactic latitude slices 
    ($ 0^\circ-10^\circ, 
      10^\circ-20^\circ, 
      20^\circ-30^\circ$ and 
     $30^\circ-40^\circ$). 
    The free-free curve is seen to lie below Foreground X 
    in all cases, typically by about an order of magnitude. 
    Foreground X therefore cannot be dominated by free-free 
    emission. The corresponding curves represent the 
    best fit models. The 10 and 15 GHz points are from the 
    Tenerife correlations, the 31.5, 53 and 90 GHz points 
    are from the DMR correlations. Upper limits are 2$-\sigma$.
    }
\end{figure*}

The goal of this {\it Letter} is to perform new cross-correlation tests 
and to present a model that reconciles these apparently contradictory 
results. Since the recently completed map from the Wisconsin H-Alpha
Mapper (WHAM) should trace free-free emission, it offers a powerful way to 
test the free-free explanation of Foreground X.
A recent cross-correlation analysis of WHAM with QMAP (de Oliveira-Costa 
\etal 2000, hereafter dOC00) failed to reveal a significant correlation, 
but this is hardly surprising since QMAP failed to detect significant 
contributions of Foreground X as well. Below we will start out in 
\S\ref{sec2} by cross-correlating WHAM with the Tenerife data, where 
Foreground X was previously detected at the $4.5\sigma$ level (dOC99). 
We will find that free-free emission is detected at levels far too low
to explain Foreground X. In \S\ref{sec3} we therefore proceed to propose 
an alternative model for what is going on.

\section{CONSTRAINTS FROM THE CROSS-CORRELATION}\label{sec2}

\subsection{Cross-correlation method}

In this section, we quantify the foreground contamination in the 
Tenerife data by cross-correlating it with a number of Galactic 
emission templates. This is done with the multi-component fit method, 
described in detail in dOC99. 
This method models the vector of Tenerife observations $\ybf$ 
as a sum
\beq{signals}
	\ybf = \Xbf\albf + \xbf_{CMB} + \nbf,
\eeq
where $\Xbf$ is a matrix whose rows contain the various foreground 
templates convolved with the Tenerife triple-beam (\ie, $\Xbf_{ij}$ 
would be the $\ith$ observation if the sky had looked like the 
$\jth$ foreground template), and $\albf$ is a vector of size $M$ 
that gives the levels at which these foreground templates are present 
in the Tenerife data.
 
The estimate of $\albf$ is computed by minimizing 
    $\chi^2 \equiv
    (\ybf - \Xbf\albf)^T {\bf C}^{-1} (\ybf - \Xbf\albf)$, 
and by modeling the covariance matrix ${\bf C}$ including both the 
experimental noise $\nbf$ and CMB sample variance in the CMB 
signal $\xbf_{CMB}$.
As in dOC99, our quoted error bars include 
the effects of both noise and chance alignments between the CMB and 
the various template maps.
The minimum-variance estimate of $\albf$ is
\beq{alpha}
   \albfHat = \left[ \Xbft \Cbf^{-1} \Xbf \right]^{-1} \Xbft \Cbf^{-1} \ybf
\eeq
with covariance matrix
\beq{varalpha}
   {\SS} \equiv \expec{\albfHat \albfHat^T} - \expec{\albfHat} \expec{\albfHat^T} =
         \left[ \Xbft \Cbf^{-1} \Xbf \right]^{-1}.
\eeq

\subsection{Data used}

We use the latest version of the Tenerife data 
(Mukherjee \etal 2000), which has more accurate offset removal 
than that used in dOC99.
To obtain accurate estimates of foreground levels, it is important
that the analysis includes all relevant emission components and the
correlations between them. 
As a synchrotron template, we use the 408~MHz survey (Haslam \etal 
1982, hereafter H82), and the 1420~MHz survey (Reich 1982; 
Reich and Reich 1986) for cross-checks.
As a template for both vibrational and spinning dust emission, 
we use the Diffuse Infrared Background Experiment (DIRBE) sky maps 
at wavelengths from 12 to 240\microm (Boggess \etal 1992).
Finally, as a tracer of free-free emission, we use the Wisconsin 
H$-$Alpha Mapper (WHAM) survey (Haffner \etal 1999).
The extent of point source contamination in the Tenerife data 
was discussed and estimated in Guti\'errez \etal (2000), and will 
therefore not be addressed in this \lletter. In practice, we just 
remove the estimated point source contribution before calculating 
the correlations. 

\begin{table}
{\footnotesize\center{\label{CorrTab} Table~1 -- Correlations for 10 and 15~GHz data.}
\begin {center}
\begin {tabular}{|l|l|rrrr|}	
\hline
 \multicolumn{1}{|c|}{$b$ \& $\nu$}	       &
 \multicolumn{1}{ c|}{Map$^{(a)}$}             &
 \multicolumn{1}{ r }{$\albfHat\pm\da^{(b)}$}  & 
 \multicolumn{1}{ r }{${\albfHat\over\da}  $}  &	 
 \multicolumn{1}{ c }{$\sigma_{Gal}        $}  &	       
 \multicolumn{1}{ c|}{$\dT$ [\mmicrok]$^{(c)}$}\\
\hline
\hline
$|b| > 20^{\circ}$	&100\microm  & 71.4$\pm$ 14.2	&{\bf  5.0}   &0.7    &  49.9$\pm$ 9.9  \\
10 GHz		  	&H82	     & 71.3$\pm$  6.3	&{\bf 11.3}   &1.0    &  71.3$\pm$ 6.3  \\
 		  	&WHAM	     & -4.7$\pm$ 19.9	&     -0.2    &0.4    &  -1.9$\pm$ 8.0  \\
\hline
                 	&100\microm  & 51.4$\pm$  8.0	&{\bf  6.4}   &0.7    &  35.9$\pm$ 5.6  \\
15 GHz		  	&H82	     &  4.1$\pm$  4.1	&      1.0    &1.0    &   4.1$\pm$ 4.1  \\
 		  	&WHAM	     &-13.6$\pm$ 10.8	&     -1.3    &0.4    &  -5.4$\pm$ 4.3  \\
\hline
$|b| > 30^{\circ}$	&100\microm  & -3.5$\pm$ 40.0	&     -0.1    &0.2    &  -0.7$\pm$ 8.0  \\
10 GHz		  	&H82	     &  5.8$\pm$  9.4	&      0.6    &0.9    &   5.2$\pm$ 8.5  \\
 		  	&WHAM	     & 39.3$\pm$ 75.2	&      0.5    &0.1    &   3.9$\pm$ 7.5  \\
\hline
                	&100\microm  & 53.7$\pm$ 26.3	&{\bf  2.0}   &0.2    &  10.7$\pm$ 5.3  \\
15 GHz		  	&H82	     & -4.5$\pm$  6.0	&     -0.8    &0.9    &  -4.0$\pm$ 5.4  \\
 		  	&WHAM	     &-38.1$\pm$ 43.6	&     -0.9    &0.1    &  -3.8$\pm$ 4.3  \\
\hline
$|b| \simgt 40^{\circ}$	&100\microm  & 65.9$\pm$ 60.5	&      1.1    &0.2    &  13.2$\pm$12.1  \\
10 GHz		  	&H82	     & -4.8$\pm$ 12.3	&     -0.4    &0.8    &  -3.8$\pm$ 9.8  \\
 		  	&WHAM	     & 19.7$\pm$107.0	&      0.2    &0.1    &   1.9$\pm$10.7  \\
\hline
                	&100\microm  & 27.0$\pm$ 43.3	&      0.6    &0.2    &   5.4$\pm$ 8.7  \\
15 GHz		  	&H82	     & -3.6$\pm$  7.5	&     -0.5    &0.8    &  -2.8$\pm$ 6.0  \\
 		  	&WHAM	     &-55.0$\pm$ 58.2	&     -0.9    &0.1    &  -5.5$\pm$ 5.8  \\
\hline
\hline
$|b| > 20^{\circ}$	&Fink12\microm &777.3$\pm$273.8  &{\bf  2.8}  & 0.03  & 23.3 $\pm$ 8.2  \\
10 GHz	          	&H82	       & 73.1$\pm$  6.3  &{\bf 11.6}  & 1.0   & 73.1 $\pm$ 6.3  \\
                  	&WHAM	       &  5.0$\pm$ 19.8  &	0.3   & 0.4   &  2.0 $\pm$ 7.9  \\
\hline
                	&Fink12\microm &763.2$\pm$169.7  &{\bf  4.5}  & 0.03  & 22.9 $\pm$ 5.1  \\
15 GHz	          	&H82	       &  5.2$\pm$  4.1  &	1.3   & 1.0   &  5.2 $\pm$ 4.1  \\
                  	&WHAM	       &-14.0$\pm$ 10.9  &     -1.3   & 0.4   & -5.6 $\pm$ 4.4  \\
\hline
\hline
$|b| > 20^{\circ}$	&Combo         &  1.0$\pm$  0.2  &{\bf  5.0}  &58.1   & 58.1 $\pm$11.7  \\
10 GHz		  	&H82	       & 68.5$\pm$  6.4  &{\bf 10.7}  & 1.0   & 68.5 $\pm$ 6.4  \\
 		  	&WHAM	       &-15.2$\pm$ 20.2  &     -0.8   & 0.4   & -6.1 $\pm$ 8.1  \\
\hline
                 	&Combo         &  1.0$\pm$  0.2  &{\bf  5.0}  &32.4   & 32.4 $\pm$ 6.5  \\
15 GHz		  	&H82	       &  4.5$\pm$  4.1  &	1.1   & 1.0   &  4.5 $\pm$ 4.1  \\
 		  	&WHAM	       &-13.9$\pm$ 10.8  &     -1.3   & 0.4   & -5.6 $\pm$-4.3  \\
\hline
\end{tabular}
\end{center}
} 
\vskip-0.1cm
\noindent{\small  
$^{(a)}$ The DIRBE, Haslam and WHAM correlations listed above correspond 
         to joint 100$\um-$H82$-$\ha fit. \\
$^{(b)}$ $\albfHat$ has units \microk (\mj)$^{-1}$ for the 100\mmicrom, 
         \mmicrok/K for the Haslam and \mmicrok/R for the WHAM template. \\ 
$^{(c)}$ $\dT \equiv (\albfHat\pm\da) \sigma_{Gal}$.}\\
\end{table}

\subsection{Cross-correlation results}

Cross-correlation results are presented in Table~1 and \Fig{dTFig}. 
All fits are done jointly for 3 templates (100$\um$ $-$ H82$-$WHAM), and 
statistically significant ($>2\sigma$) correlations listed in Table~1 
are in boldface. 
Since the fluctuation levels depend strongly on Galactic latitude,
we perform our analysis for six different latitude slices of roughly 
equal area (\Fig{dTFig}) as well as for three different latitude cuts 
(Table~1): $20^{\circ}, 30^{\circ}$ and the {\it Tenerife cut}  
(which consists of data inside the region $160^{\circ} < {\rm RA} < 250^{\circ}$, 
corresponding to Galactic latitudes $|b| \simgt 40^{\circ}$).
For definiteness, we use the DIRBE 100$\um$ channel when placing all
limits shown in this subsection since it is the least noisy of the 
DIRBE channels, and the Haslam map since it is the \syn template 
at lowest frequency.

\begin{table}
{\footnotesize\center{\label{DustHaTab} Table~2 -- 100$\um-$\ha correlations.}
\begin {center}
\begin {tabular}{|l|l|c|c|}
\hline
 \multicolumn{1}{|c|}{Authors$^{(a)}$}       &
 \multicolumn{1}{c|} {$b$}	             &
 \multicolumn{1}{c|} {$\albfHat\pm\da^{(b)}$}& 
 \multicolumn{1}{c|} {${\albfHat\over\da}$} \\
\hline
\hline
McCullough (1997)    & $b=-65^\circ$			     &0.79 $\pm$0.24  &{\bf  3.3} \\
Kogut (1997)         & $b=-21^\circ$			     &0.85 $\pm$0.44  &      1.9  \\
                     & $b=+27^\circ$			     &0.34 $\pm$0.33  &      1.0  \\
dOC00                & $b=+27^\circ$ 		     	     &0.25 $\pm$0.19  &      1.3  \\
\hline
This work$^{(c)}$    & $|b| > 20^\circ$                      &0.18 $\pm$0.07  &{\bf  2.6} \\
                     & $|b| > 30^\circ$                      &0.17 $\pm$0.04  &{\bf  4.3} \\
                     & $Tenerife~Cut$                        &0.04 $\pm$0.03  &      1.3  \\
                     &$00^\circ \simlt |b| \simlt 10^\circ$  &0.004$\pm$0.100 &      0.04 \\
                     &$10^\circ \simlt |b| \simlt 20^\circ$  &0.04 $\pm$0.17  &      0.2  \\
                     &$20^\circ \simlt |b| \simlt 30^\circ$  &0.16 $\pm$0.15  &      1.1  \\
                     &$30^\circ \simlt |b| \simlt 42^\circ$  &0.16 $\pm$0.03  &{\bf  5.3} \\
                     &$42^\circ \simlt |b| \simlt 56^\circ$  &0.14 $\pm$0.10  &      1.4  \\
                     &$56^\circ \simlt |b| \simlt 90^\circ$  &0.14 $\pm$0.08  &      1.8  \\
\hline
\end{tabular} 
\end{center}
}
\vskip-0.1cm
\noindent{\small  
$^{(a)}$ Patches are centered in the indicated $b$ coordinates. \\
$^{(b)}$ $\albfHat$ has units R (\mj)$^{-1}$. \\ 
$^{(c)}$ Our results are for the 15~GHz data. \\
}
\end{table}

\Fig{dTFig} shows the corresponding fluctuations in antenna 
temperature in the Tenerife data ($\delta T = \albfHat \sigma_{Gal}$, 
where $\sigma_{Gal}$ is the standard deviation in the template map).
The Haslam and DIRBE correlations are seen to be consistent with those
from dOC99. Synchrotron emission (triangles) generally dominates 
the rms foreground fluctuations at 10 GHz. At 15 GHz, on the other hand, 
Foreground X (squares) is seen to dominate except in the Galactic plane 
itself. The key new result here is the inclusion of the WHAM data, showing 
that free-free emission (circles) is about an order of magnitude below 
Foreground X over the entire range of frequencies and latitudes where it 
is detected. This means that Foreground X cannot be explained as
free-free emission.
The corresponding values of $a_i$ (in units of $\mu$K/R)
are consistent with gas at 8000K (Bennett \etal 1992) for 
latitudes $|b|=0^\circ-10^\circ$ and frequencies $\ge$ 30 GHz, with 
substantial scatter elsewhere\footnote{
	Although the cross-correlation technique can accurately 
	determine the dominant components, the detailed results for 
	strongly subdominant components must be taken with a grain of salt.
	This is because any complications with the dominant components
	(\eg, slight spatial variations in their frequency dependence) 
	which are not included in the model of \eq{signals} 
	~will act as 
	excess noise on the remaining components. Although the analysis
	clearly demonstrates that free-free emission is subdominant,
	the formal error bars on this component are therefore likely 
	to be smaller than the true errors.
	}.

As in de Oliveira-Costa \etal (1997; 1998), dOC99 and dOC00, the 
cross-correlation software was tested by analyzing constrained 
realizations of CMB and Tenerife instrument noise. From 1000 
realizations, we recovered unbiased estimates $\albfHat$ with a 
variance in excellent agreement with equation (\ref{varalpha}). 
As an additional test, we computed 
     $\chi^2 \equiv 
     (\ybf - \Xbf\albf)^T {\bf C}^{-1} (\ybf - \Xbf\albf)$
and obtained $\chi^2/N \approx 1$ in almost all cases.

We also performed a joint fit retaining the 
100$\um$, Haslam and WHAM templates in the Tenerife observing region,
but replacing the Tenerife data by COBE DMR data at 31.5, 53 and 90~GHz. 
These results are also plotted in \Fig{dTFig} by multiplying the measured 
coefficients $\albfHat$ by template rms $\sigma_{Gal}$ corresponding 
to the Tenerife triple beam, and show that our conclusions extend to 
higher frequencies as well\footnote{
	A full DMR$-$WHAM correlation analysis is in progress 
	(Kogut \etal 2002), so the results presented here should be viewed 
	as preliminary. For all-sky cross-correlations, large scale 
	variations in the WHAM map will need to be better accounted for.
	}.

Since we find substantially lower levels of free-free emission than 
Foreground X, the corresponding templates  (100\microm and WHAM) cannot 
be very highly correlated. We confirmed this by a direct cross-correlation 
analysis between these two templates, as seen though the Tenerife 
triple beam in the Tenerife observing region, and found the dimensionless 
correlation coefficient between the two maps to be in the range 5\%$-$30\% 
depending on Galactic latitude. The corresponding results in physical units 
(R/MJy sr$^{-1}$) are shown in Table~2.
Since the statistical properties of these maps are not accurately
known, we estimated the error bars by repeating the analysis with
one of the templates replaced by $4\times 36=144$ transformed 
maps, rotated around the Galactic axis by multiples of 10$^\circ$ 
and/or flipped vertically and/or horizontally. 
Significant correlations are found only for the $b>20^\circ$ and 
$b>30^\circ$ cuts, and the bulk of this correlation seems to come 
from a region of the sky between $30^\circ<b<42^\circ$.
 
For comparison, see McCullough (1997), Kogut (1997) 
and dOC00 results shown in Table~2.
This table indicates significant variations across the sky in
the relationship between \ha and the 100\microm emission, 
which could be related to variations in Hydrogen ionization fraction.

Finally, writing the frequency dependence as $a\propto\nu^{\beta}$
and recalling that the correlation coefficient is, by definition, 
$a=1$K $/\mu$K$=10^6$ for H82  at 408  MHz, we obtain the spectral 
index limits of
      $-2.9 \simlt \beta \simlt -3.6$ for the 10~GHz$-$H82 correlation and 
      $-3.2 \simlt \beta \simlt -3.7$ for the 15~GHz$-$H82 correlation.
These values are slightly steeper than the canonical sub-GHz slope of 
      $-2.7 \simlt \beta \simlt -2.9$
(Davies \etal 1998; Platania \etal 1998), but
consistent with a steepening of the spectrum of cosmic ray electrons 
at higher energies (Rybicki and Lightman 1979). The relatively high 
Tenerife synchrotron signal seen in Table~1 could be interpreted 
as slight spatial variability of the frequency dependence 
(Tegmark \etal 2000).

\subsection{Power spectrum of foregrounds}

\begin{figure}[tb]
\vskip+0.1cm
\centerline{{\vbox{\epsfxsize=9.0cm\epsfysize=6.5cm\epsfbox{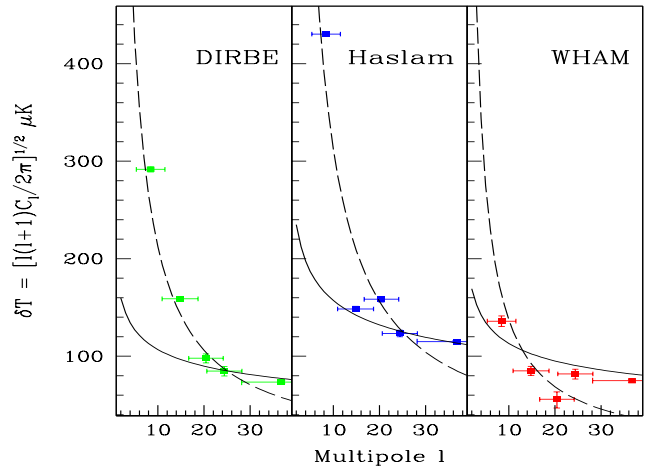}}}}
\vskip-0.1cm
\figcaption{\label{powerspectrum}
    Angular power spectra for the DIRBE, Haslam
    and WHAM components as seen in the Tenerife data. 
    Results are from the $20^\circ-30^\circ$ slice shown in 
    \Fig{dTFig}. The curves show $\l^{-6}$ (dashed line) 
    and $\l^{-3}$ (solid line) power laws. 
    }
\end{figure}

\Fig{dTFig} shows the rms contribution of each foreground.
In order to understand which angular scales contribute most
to this rms, we compute the angular power spectra of the 
template emissions.
The angular power spectra of the DIRBE, Haslam
and WHAM components are shown in \fig{powerspectrum} for 
the $20^\circ-30^\circ$ slice of \fig{dTFig}, scaled to 15~GHz 
using the correlation coefficients found above.
All three power spectra are seen to be compatible with a power 
laws of $\l^{-3}$ (solid line) at small angular scales 
($\l>10-15$)\footnote{
	Our DIRBE result is compatible with that reported
	by Wright (1998).
	},
while DIRBE and Haslam are seen to steepen further on very 
large scales ($\l<10-15$) to a power law closer to 
$\l^{-6}$ (dashed line).

We remind the reader that there is, strictly speaking, no such 
thing as the power spectrum of a Galactic foreground, since 
the latitude dependence implies that it is not an isotropic 
random field.
What we refer to here as the power spectrum of a foreground  
is simply the quantity that we care about in practice: its 
contribution to the measurement of a CMB power spectrum.
We have therefore computed the curves in \fig{powerspectrum}
by convolving the template maps with the Tenerife triple beam 
and treated the result as if it were a CMB map, computing 
quadratic band power estimators as described in Tegmark (1997)
and Bond, Jaffe and Knox (1998).

\section{AN ALTERNATIVE MODEL FOR FOREGROUND X}\label{sec3}

The results presented above show that free-free emission cannot 
explain Foreground X. Since there is presently only one other strong 
contender, spinning dust as proposed by DL98, 
the case would appear to be closed\footnote{
	There is also the possibility of Foreground X is due to 
	magnetic dipole emission from ferromagnetic grain materials
	(Draine and Lazarian 1999). This model can be ruled 
	out if foreground X can be shown to correlate better
	with 12\microm than 100\microm emission.}. 
However, as we describe below, the situation seems to be a little 
more complicated.

\begin{figure*}[tb]
\centerline{{\vbox{\epsfxsize=18.2cm\epsfbox{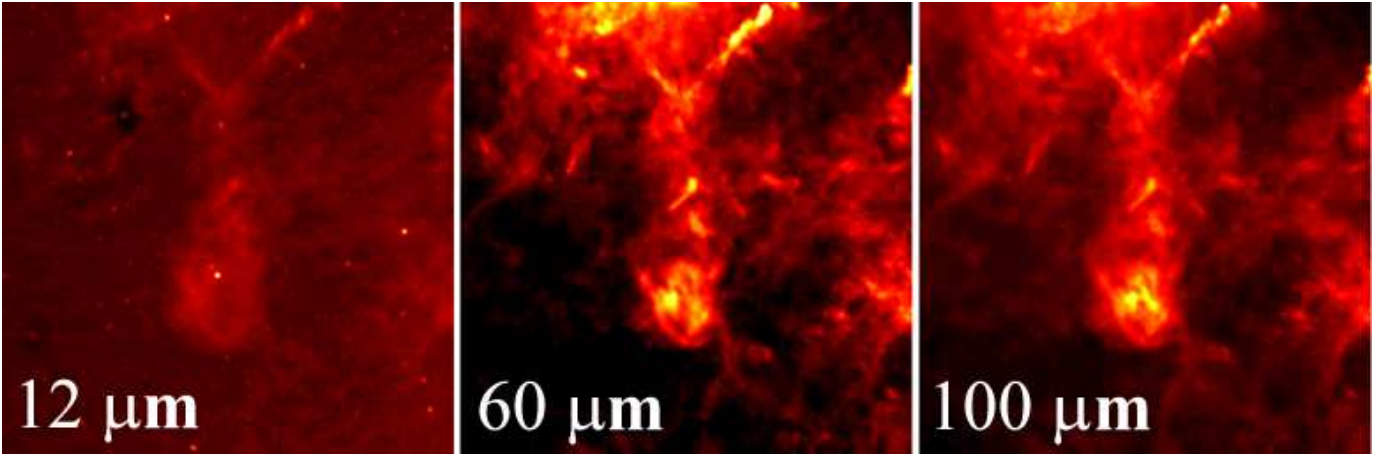}}}}
\figcaption{\label{mushroom}
    12, 60 and 100\microm IRAS images from the ``mushroom'' cloud MBM 20. 
    These images are squares of 12.8$^{\circ} \times $12.8$^{\circ}$ 
    centered at $l=210.9, b=-34.5$. A simple visual
    comparison of these IRAS images at different frequencies 
    suggests that although the large-scale features generally 
    match up, small scale features can be quite different. 
    Experiments such as ground-based interferometers, MAP and Planck LFI may 
    have the angular resolution at the relevant low frequecies 
    that are needed to be able to confirm or to rule out our grain 
    segregation model.    
    }
\end{figure*}

100\microm correlations have been detected at high significance on
large ($3^\circ-7^\circ$) angular scales:
	 at $4.2\sigma$ at 31 GHz by DMR, 
	 at $6.6\sigma$ in the 19 GHz survey, 
	 at $6.4\sigma$ at 15 GHz by Tenerife and 
	 at $5.1\sigma$ at 10 GHz by Tenerife. 
These detections reveal a spectrum rising towards lower frequencies, 
with hints of a plateau or turnover between 10 and 15 GHz.
On intermediate (degree) scales, the Saskatoon data gave a marginal 
detection while the QMAP data gave only an upper limit.
At small (arcminute) scales the situation is even more confusing:
although F01 recently reported a correlation of an HII region 
(LPH 201.663+1.643) and a dark cloud (L 1622) at 5, 8 and 10~GHz with 
the SFD98 100 $\mu$ map (Schlegel \etal 1998), these values are not 
in concordance with originally reported value given by OVRO at 14.5~GHz 
(Leitch \etal 1997)\footnote{
	The statistical significance of this result is unclear, 
	since the OVRO measurement was published without error bars.
	}.
Moreover, only the dark cloud correlation detected by F01 seems to be 
clearly consistent with our (large scale) Tenerife result of $\albfHat =$ 
71.4$\pm$ 14.2 \mmicrok (\mj)$^{-1}$ (see Table~1).
This discrepancy cannot be blamed on variations in dust 
column density between different sky patches, since they
would not affect the dust-to-CMB conversion factor $\albfHat$.
In summary, in average the most solid detections of Foreground X are 
all on scales of a few degrees or larger. Why is this foreground so 
elusive on smaller scales?

\subsection{A grain size segregation model}

Here we will argue in support of a model where Foreground X is spinning 
dust after all, and the problem is with the template used to 
find it, \ie, 100\microm dust emission. 

Galactic dust grains come in a wide range of sizes 
(Weingartner and Draine 2000a).
The 100\microm template is dominated by large dust grains that are close
to thermal equilibrium with the interstellar radiation field, 
and are radiating thermally at temperatures around 20K (Reach \etal 1995).
In contrast, the small grains that can spin fast enough to produce 
the rotational emission of Draine and Lazarian cool faster than the mean 
time between two photon absorption events. Therefore, they spend most 
of their time near their ground state. They get heated to as much as a few 
hundred Kelvin when they absorb a photon, and radiate non-thermally 
with a spectrum that is both bluer and broader than for the large 
thermalized grains (Draine and Lee 1984).
As a result, the emission from small grains peaks at shorter 
wavelengths like 10$-$30\mmicrom. Indeed, it is such non-equilibrium 
behavior that enables the Draine and Lazarian emission mechanism to work.
 
Previous work has generally assumed that the relative abundance 
of grain sizes is independent of position and that the 100\microm
map (dominated by large grains) is a good tracer for small grains 
as well. However, this seems to not be the case. A simple visual
comparison of the IRAS maps at different frequencies suggests that 
although the large-scale features generally match up, small scale 
features can be quite different (see \fig{mushroom}).

Our proposed solution to the spinning dust puzzle is therefore that 
this component (small grains) only correlates well with 100\microm
emission (large grains) on fairly large angular scales. Occasional 
agreements on small scales would of course not be precluded (which 
could explain the F01 and Leitch \etal 1997 results), but should not be 
expected to hold in general.

As pointed out by Weingartner and Draine (2000b),
it is not physically implausible for small and large dust grains 
to be separated. They showed that small dust grains can
stay locked to the gas, while large dust grains in diffuse 
clouds can drift due to the effects of anisotropic starlight. 
After 10$^7$ yr, a typical lifetime of a diffuse interstellar cloud, 
such drifts can separate large and small grains by up to 
a few degrees at high Galactic latitudes.

\subsection{Testing our model}

\begin{figure}[tb]
\vskip-1.6cm
\centerline{\hglue-5mm{\vbox{\epsfxsize=10.0cm\epsfbox{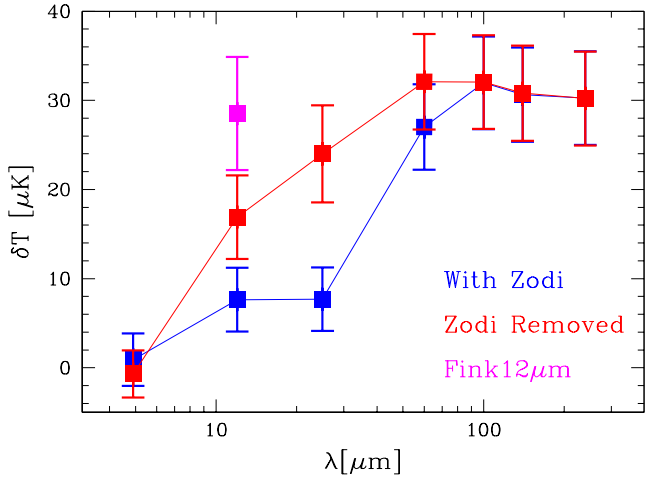}}}}
\vskip-1.6cm
\figcaption{\label{Dirbes}
    Dust correlations for the 12$-$240\microm DIRBE maps at 15~GHz.
    Correlations of DIRBE maps with zodi are shown in blue, 
    correlations of DIRBE maps with zodi removed are shown in
    red, and correlation with 12\microm map with zodi and point
    source removed is shown in magenta. 
    }
\end{figure}

A first prediction of this model is that shorter wavelength 
dust maps should trace spinning dust at least as well as a 
100\microm map (indeed, better on small scales)\footnote{ 
	This prediction was also made by DL98, who argued that 
	the 30~GHz emission 
	and diffuse 12\microm emission should be correlated, 
	since both originate in grains containing $\approx$
	100 atoms.}.
To test this, we repeated our analysis with the 100\microm map 
replaced by DIRBE maps at other wavelengths. The results are shown 
in \fig{Dirbes} (lower curve). A first glance, it appears inconsistent 
with our prediction. However, contributions other than Galactic 
dust can clearly spoil the correlation, and several such 
contaminants are known to be present. Indeed, the figure shows 
that removing zodiacal emission from dust in our solar system 
increases the short wavelength correlations so much that the 
60\microm map traces spinning dust marginally better than the 
100\microm map (upper curve). The shorter wavelength maps also 
contain a substantial point source contribution. We used a new 
merged and destriped DIRBE+IRAS 12\microm map\footnote{
	 This new map was constructed in the same fashion as
	 SFD98 100\microm map (for details about how such composite
	 maps are made see Schlegel \etal (1998) or 
	 {\it http://www.astro.princeton.edu/$\sim$schlegel/dust/index.html}).
	 }
to eliminate all 5$-\sigma$ point source before convolving it 
with the Tenerife beam.
\Fig{Dirbes} shows that this new 12\microm template traces spinning 
dust almost as well as the 100\microm map (``Fink12\mmicrom''). 
For comparison, correlations with the 12\microm map are shown 
in Table~1, bottom.

We tried two additional approaches to further increases the 
correlations (none of which helped more than marginally).
The first was to apply a zodiacal cut on the zodi removed maps. 
The fact that this failed to increase correlations suggests that 
the zodi-removal performed by the DIRBE team was already 
adequate for our purposes.
The second was to compute the linear combination of all DIRBE 
maps that gave the best correlation, to see if contaminants 
could be indentified and subtracted spectrally. Correlations 
with this composite map are also shown in Table~1, bottom (see ``Combo'').
Our interpretation of these negative results is that our large
dust grain templates are already tracing spinning dust quite 
well on large scales, so that it is impossible to do much 
better. Rather, the key tests will involve the correlation
on small angular scales.

\section{DISCUSSION}

We have used the WHAM map to show that DIRBE-correlated 
microwave emission (Foreground X) cannot be explained as 
free-free emission. Since the spinning dust alternative has been 
challenged by recent small-scale observations, we have argued for 
an alternative model involving small scale grain segregation, 
where small and large dust grains are well correlated only on 
large angular scales.

We found that the zodi-cleaned 60\microm DIRBE map traces Foreground X
marginally better than the 100\microm map, and that even the 12\microm 
map does a good job after zodi and point source removal. Although these 
preliminary results are mildly encouraging, the smoking-gun test 
of this model will involve cross-correlating dust maps at various 
frequencies with high-resolution CMB data in the 10-30~GHz range.

Fortunately, such maps should be available shortly, from CMB 
experiments such as ground--based interferometers,  
the NASA MAP satellite\footnote{http://map.gsfc.nasa.gov/} and 
the Planck LFI\footnote{http://astro.estec.esa.nl/SA-general/Projects/Planck/}. 
For instance, the expected sensitivity of the MAP 22~GHz channel 
is about 11\microk per 56 arcminute pixel. For comparison, the expected 
small-scale fluctuations from spinning dust at 22~GHz are of order 20\microk 
at high Galactic latitudes, ranging from a few to $10^2$\microk 
from a clean to a dirty 
region\footnote{We made this estimate by high-pass filtering the 
		Fink12\microm map to the degree scale $-$ this was 
		done by taking the difference of two Fink12\microm
		maps, one smoothed by $0.7^\circ$ and another by $2^\circ$.
		Using $\albfHat\approx$760\mmicrom/(\mj) 
		from Table~1, we converted this difference map to 
		\microk, and finally made a small spectral correction 
		from 15 to 22~GHz.}.
Since MAP will measure tens of thousands of such pixels, 
it should be readily able to confirm or rule out our model.
The prediction is that the small-scale signal will be 
substantially better traced by shorter wavelength dust maps.
Since various contaminants may be important in these short wavelength
maps (\eg, polycyclic aromatic hydrocarbons $-$ PAHs), it will also be
worth performing a multicomponent fit using dust maps at all available 
frequencies, to find the linear combination of the dust maps that 
constitutes the best spinning dust template.
High-resolution low frequency ground-based experiments such as 
CBI\footnote{http://astro.caltech.edu/$\sim$tjp/CBI/} and 
DASI\footnote{http://astro.uchicago.edu/dasi/} 
may also be able to test our hypothesis, but this is far from 
clear because they operate in Ka-band (26-36 GHz), where the 
spinning dust signal is small.
%
%
In conclusion, we have proposed a resolution to the puzzle
of Foreground X. Observations during the coming year should 
be able to test it.

\bigskip

We would like to thank Bruce Draine for useful comments. 
Support was provided by the NASA grant NAG5-9194
({\it n\'e} NAG5-6034) and the University of Pennsylvania 
Research Foundation. 
WHAM is supported by the NSF through grant AST96-19424.
We acknowledge the NASA office of Space Sciences, the COBE flight team, 
and all those who helped process and analyze the DIRBE data.


\clearpage

\end{document}